\documentclass[usenatbib]{mn2e}
\usepackage{graphicx}

\title[Infall and SiO emission in V838 Mon]{Infall and SiO emission in V838 Mon}
\author[Rushton et al.]
 {M. T. Rushton$^{1}$, 
 T. R. Geballe$^{2}$,
 A. Evans$^{1}$,
 B. Smalley$^{1}$, 
 J. Th. van Loon$^{1}$,
\newauthor S. P. S. Eyres$^{3}$ \\ 
$^1$ Astrophysics Group, School of Physical and Geograhical Sciences, Keele University, Keele, Staffordshire, 
ST5 5BG, UK \\
$^2$ Gemini Observatory, 670 N. A`oh\={o}k\={u} Place, University Park, Hilo, HI 96720, USA \\
$ 3$ Centre for Astrophysics, University of Central Lancashire, Preston, PR1 2HE, UK}

\date{Version of 22 Oct 2004}

\pagerange{\pageref{firstpage}--\pageref{lastpage}}

\def\LaTeX{L\kern-.36em\raise.3ex\hbox{a}\kern-.15em
    T\kern-.1667em\lower.7ex\hbox{E}\kern-.125emX}

\newcommand{\kms}{\mbox{\,km~s$^{-1}$}}

\newcommand{\pion}[2]{{#1}\,{\sc {#2}}}

\newcommand{\ltsimeq}{\raisebox{-0.6ex}{$\,\stackrel 
           {\raisebox{-.2ex}{$\textstyle <$}}{\sim}\,$}} 
\newcommand{\gtsimeq}{\raisebox{-0.6ex}{$\,\stackrel
           {\raisebox{-.2ex}{$\textstyle >$}}{\sim}\,$}}

\begin{document}
\label{firstpage}
\maketitle
\begin{abstract}
We present moderate and high resolution infrared spectroscopy of the peculiar eruptive
variable V838 Mon, which underwent a series of remarkable outbursts in early 2002. During the
period covered by our observations, 2002 December--2003 December, the near-infrared spectrum
continued to show many of the characteristics of a very cool supergiant. However, throughout
this period the spectrum also revealed strong and variable SiO first overtone emission, and
Pa$\beta$ emission. The  2003 December spectrum contained a series of \pion{Ti}{i} lines with
inverse P Cygni profiles. This is clear evidence that some material is falling inward towards
the star.   
\end{abstract}

\section{Introduction}

The multiple outburst episode of V838 Mon in early 2002 has been well documented
\citep{munari,kimeswenger,crause,banerjee,wisniewski,rushtona}. The object was first detected
in outburst on 2002 January 6 ($V_{\rm max}=10$) \citep{brown}, and subsequent outbursts
developed in 2002 February ($V_{\rm max}=6.7$) and 2002 March ($V_{\rm max}=7$). Optical
spectroscopy showed a cool, reddened object, whose continuum shape and absorption spectrum
were broadly consistent with those of a K supergiant. However, many of the spectral lines
displayed P Cygni profiles, indicating outflow speeds of $\sim$200\kms, and  exhibited strong
variability. The post-eruption phase was characterised by a dramatic increase in $V-I$,
accompanied by a rapid trend towards later spectral types. As of 2002 May the visual
magnitude had faded by $\sim7$ mag since the 2002 March peak, and the effective temperature
was 2400~K \citep{banerjee}. By 2002 September further bolometric fading had revealed the
presence of a possible B3~V companion \citep{desidera}. At this time the near-infrared
spectrum of V838 Mon displayed very deep H$_{2}$O, CO, TiO, VO and AlO bands, and
superficially resembled L-type brown dwarfs, indicating a very cool giant/supergiant
\citep{evans}. 

V838 Mon may not be a unique object. The similarities with V4332 Sgr and M31~RV are
broad-based \citep{munari,kimeswenger,boschi}, and a further analogue may be found in the
``Peculiar Variable in Crux'' \citep{della}. Recently \citet{banerjeeb} presented a
near-infrared spectrum of V4332 Sgr, which also displayed strong AlO bands, but in emission.
Accretion models involving stellar mergers \citep{soker} and planets \citep{retter} have been
proposed to collectively explain this group of unusual objects.

Here we present some of our latest observations of the near-infrared monitoring program of
V838 Mon, emphasizing the discovery of SiO overtone emission and inverse P-Cygni profiles on
some atomic lines.  

\section{Observations}

\begin{table*}
\caption{Observing log.}
\label{log}
\begin{tabular}{@{}lcccccc}
\hline
\multicolumn{1}{c}{Observation date}&
\multicolumn{1}{l}{Wavelength coverage}&
\multicolumn{1}{c}{Resolving Power}&
\multicolumn{1}{c}{Std. Star}\\
\multicolumn{1}{c}{(UT)}&
\multicolumn{1}{c}{($\mu$m)}&
\multicolumn{1}{c}{}&
\multicolumn{1}{c}{}\\
\hline
\multicolumn{1}{l}{2002 December 17}&
\multicolumn{1}{c}{2.9--4.1}&
\multicolumn{1}{c}{1400}&
\multicolumn{1}{c}{BS2530 (F2~V)}\\
\multicolumn{1}{l}{2003 April 4}&
\multicolumn{1}{c}{2.9--4.1}&
\multicolumn{1}{c}{1400}&
\multicolumn{1}{c}{BS2530 (F2~V)}\\

\multicolumn{1}{l}{2003 October 6}&
\multicolumn{1}{c}{2.9--4.1}&
\multicolumn{1}{c}{1400}&
\multicolumn{1}{c}{BS2530 (F2~V)}\\

\multicolumn{1}{l}{2003 December 6}&
\multicolumn{1}{c}{1.278--1.286}&
\multicolumn{1}{c}{20,000}&
\multicolumn{1}{c}{BS1879 (O8 III)}\\

\multicolumn{1}{l}{2003 December 6}&
\multicolumn{1}{c}{3.989--4.021}&
\multicolumn{1}{c}{20,000}&
\multicolumn{1}{c}{BS1879 (O8 III)}\\

\multicolumn{1}{l}{2003 December 17}&
\multicolumn{1}{c}{4.016--4.046}&
\multicolumn{1}{c}{20,000}&
\multicolumn{1}{c}{BS1879 (O8 III)}\\

\hline
\end{tabular}
\end{table*}

Low and high resolution infrared spectroscopy of V838 Mon were obtained on
numerous occasions between late 2002 and end 2003, using the facility
instrument CGS4 \citep{mountain} on the United Kingdom Infrared Telescope
(UKIRT). An observing log for the specific observations presented here from
that time interval is provided in Table~\ref{log}; additional spectra obtained
during that period will be reported later. The observations employed the CGS4
40~l/mm and echelle gratings at resolving powers of $\sim$1400 and 20,000
respectively. Flux calibration and removal of telluric absorption features were
achieved by dividing the target spectra by spectra of nearby calibration stars.
To avoid spurious emission lines in the ratioed spectra due to hydrogen
absorption lines in the calibration stars, the Pa$\beta$, Pf$\gamma$ and
Br$\alpha$ lines were removed from the calibration stars by interpolation prior
to ratioing. The low resolution ratioed spectra were then multiplied by a
blackbody corresponding to the effective temperature and broadband photometric
magnitudes of the standard star to give the final flux-calibrated spectra. The
high resolution spectra were scaled to make the continuum equal to unity after
division. At low resolution, wavelength calibration was carried out by
obtaining argon lamp spectra immediately before the target and is accurate to
better than 0.001~$\mu$m. Argon lines also were used to calibrate the echelle
spectrum at 1.28~$\mu$m, but telluric N$_{2}$O absorption lines in the spectrum
of the calibration star were used to calibrate the 4~$\mu$m echelle spectra, in
both cases to an accuracy of $\pm$3\kms.

\section{The OH and S\lowercase{i}O Bands}

\begin{figure*}
\centering
  \includegraphics[angle=0, width=12cm]{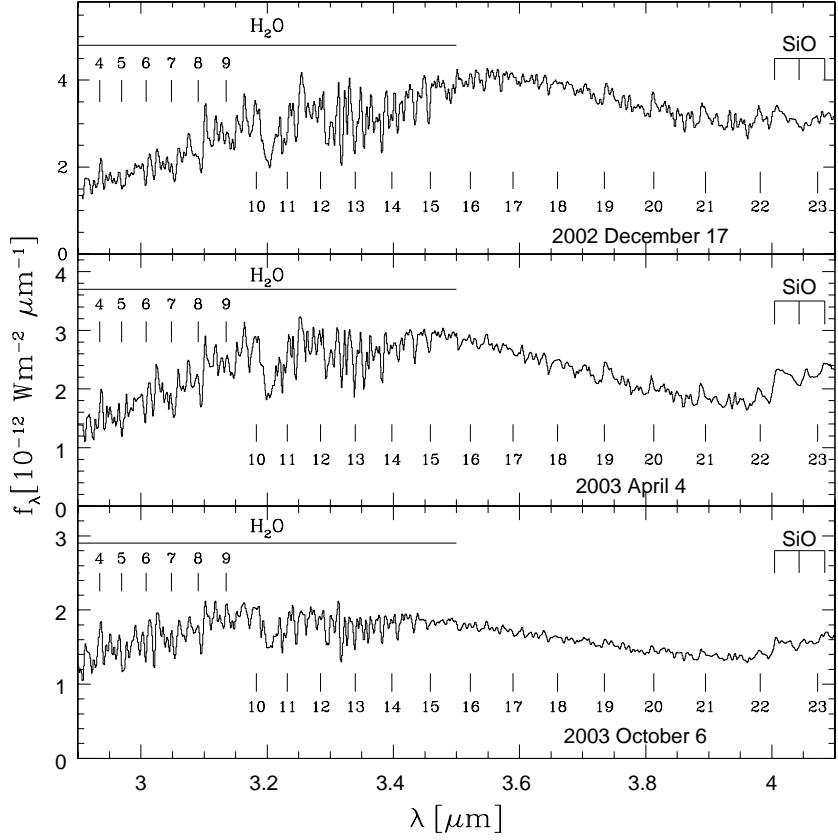}
  \caption{$LL'$ band spectra of V838 Mon on the dates indicated. The
vertical lines denote the positions of the OH fundamental transitions
shown with their corresponding $J$ values.
The wavelengths of the SiO $\upsilon=2\rightarrow0$, $3\rightarrow1$ and
$4\rightarrow2$ bandheads are also shown.}
\label{lband}
\end{figure*}

The spectral evolution of V838 Mon (2002 December$-$2003 October) in the range
2.9--4.1~$\mu$m is shown in Figure~\ref{lband}. The bulk of the detail in the 2.9--3.5~$\mu$m
region is due to a combination of the $\upsilon_1$ and $\upsilon_3$ stretching bands of
H$_{2}$O. We have observed strong water bands in V838 Mon previously in $JHK$ spectroscopy,
as discussed in \citet{evans}.

The OH fundamental ($\Delta\upsilon=1$) $P$ branch lines straddle this
wavelength range in late-type, O-rich stars. The line positions of the
$\upsilon=1\rightarrow0$ band, obtained from \citet{maillard} are indicated in
the Figure. The line strengths are usually slowly varying across the band, as
the excitation energy varies slowly as a function of $J$. The $P(19)-P(23)$
lines are the only OH lines present in the spectra and their intensities
clearly fade over the period of our observations. The lack of higher excitation
OH emission lines can be ascribed to temperature, but it is difficult to
account for features at lower $J$ unless OH absorption, increasing towards
lower $J$, is also present. Higher resolution observations might reveal such
absorption.  OH transitions involving higher vibrational states do not appear to
contribute significantly in this wavelength range.

The SiO  first overtone ($\Delta \upsilon=2$) is conspicuous in the
$\lambda\ga4~\mu$m region in emission, as was reported by \citet{lynch}.
In all of our spectra the $\upsilon = 2\rightarrow0$ band is clearly
visible, but there is little or no evidence for the $\upsilon =
3\rightarrow1$ bandhead. Therefore only the lowest vibrational states
$\upsilon\ltsimeq2$ are significantly populated.

To produce a significant population in the $\upsilon=2$ level by collisional excitation from
$\upsilon=1$ the density must be $\ga$A$_{10}$/C$_{12}$, where A$_{10}$ is the Einstein
coefficient for spontaneous decay from $\upsilon=1$ to $\upsilon=0$. For SiO-H$_2$ collisions
we can estimate the rate constant C$_{12}$ for collisional excitation from the $\upsilon=1$
to the $\upsilon=2$ level using the vibrational relaxation time formula given in
\citet{millikan}. Assuming, as is usual \citep{scoville}, the de-excitation cross sections
scale like the radiative matrix elements, and using detailed balance, we obtain
C$_{12}\simeq5.6\times10^{-13}$~cm$^{-3}$. The required density is then $n_{\rm
H_2}\gtsimeq9.2\times10^{12}$~cm$^{-3}$ at the excitation temperature of the SiO (see below).
This will be an overestimate if there is a significant fractional amount of atomic hydrogen
($n_{\rm H}\gtsimeq n_{\rm H_2}/100$), since laboratory experiments with CO show that the
rate coefficents for collisions with H are $\sim100$ times larger than those involving H$_2$
(see \citealt{scoville} and references therein).

\begin{figure*}
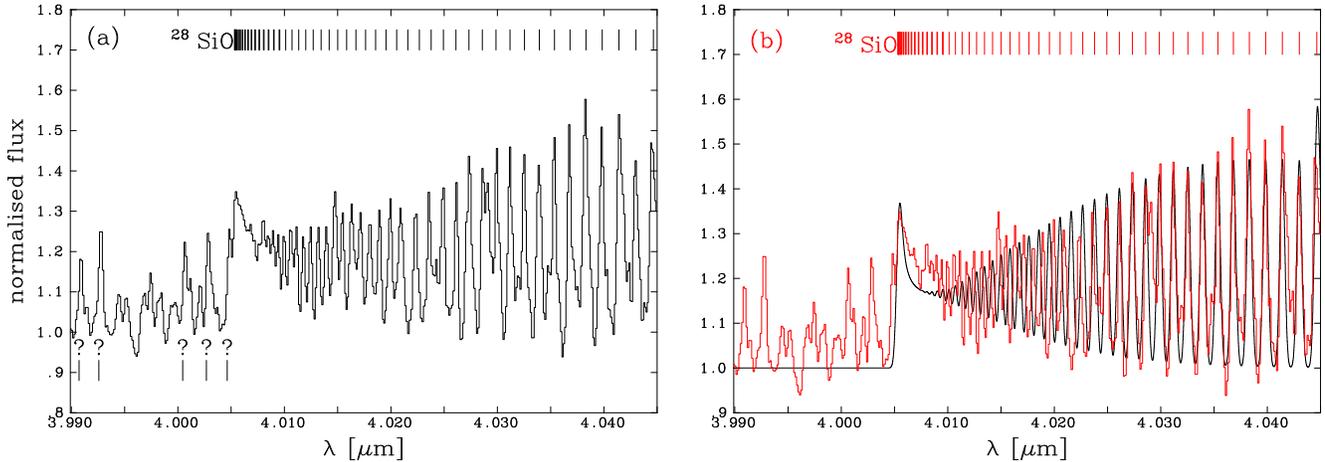

\setlength{\unitlength}{1cm}
\begin{picture}(12.0,8)
\put(-0.3,-4.0){\includegraphics{mtr2a.eps}}
\put(-0.3,-4.0){\includegraphics{mtr2b.eps}}
\end{picture}
\caption{(a) 2003 December high resolution spectrum of V838 Mon showing the
SiO first
overtone
emission. The figure also shows five unidentified emission lines
. (b) Best fitting model spectrum computed with an excitation temperature of $T=1200$~K,
superposed on the observed spectrum. The positions of the $^{28}$SiO $\upsilon=2\rightarrow0$ 
$R$ branch lines, redshifted so they occur at $+79$\kms\, (heliocentric),
are indicated in each panel (see text for details). }
\label{highressio}
\end{figure*}

Figure~\ref{highressio} shows a later spectrum at high resolution in the
spectral range 3.98--4.05~$\mu$m. This covers portions of the $\upsilon =
2\rightarrow0$ band of the main isotopomers $^{28}$Si$^{16}$O and
$^{29}$Si$^{16}$O. The fine structure is now resolved and emission in the
individual rotational-vibrational lines is clearly apparent; all of the SiO
lines in this range are $R$ branch transitions. The $^{28}$SiO line positions
used in Figure~\ref{highressio} were obtained from \citet{lovas}; these include
the positions of the $R$ banch lines extending from $R$(19) at 4.0452~$\mu$m to
the bandhead at $R$(68) (4.0043~$\mu$m). The locations of lines from higher $J$
values are not shown; the wavelengths of these overlap those of the ascending
$R$ branch but the lines are undoubtably much weaker. The radial
velocity of the low $J$ lines is $+79\pm4$\kms\,(heliocentric).

In Figure~\ref{highressio} we present a preliminary attempt to model the SiO
emission, using a simple, optically thin and isothermal slab with a Boltzmann
population. The transition probabilities of the $^{28}$SiO lines used in the
model were calculated using the dipole matrix elements given in
\citet{tipping}. The less-abundant $^{29}$SiO isotopic species is not included
in the model. However, there is no clear evidence for its detection in this
complex spectrum. The $^{29}$SiO bandhead occurs at 4.029~$\mu$m, and the
terrestrial abundance ratio $^{28}{\rm Si}/^{29}{\rm Si}$ is $\sim20$. The lack
of isolated lines in the spectrum makes it difficult to yield a definitive
isotopic ratio, and it is only reasonable to conclude $^{28}{\rm Si}/^{29}{\rm
Si}>5$ from the present data.

We find that a model with SiO excitation temperatures of $T\sim1200$~K is the
best representation of the observed spectrum (a more detailed analysis will be
presented elsewhere). This is consistent with the low temperatures seen
elsewhere in the infrared spectrum \citep{evans}, and suggests the SiO must be
located close to the stellar photosphere, rather than further out in the
extended atmosphere. However, a lower SiO excitation temperature of 790\,K
was found by \citet{lynch} from their earlier spectra (2003 January and
February). Our low resolution data clearly show the SiO emission varying over
the twelve months to 2003 December (see Figure~\ref{highressio}), with the
strongest emission in the 2003 April spectrum, and it is just as pronounced in
the Lynch et al. spectra, albeit at a lower resolution (see their Figure\,8). A comparison of their results
with our 2002 December 17 spectrum, suggests that the flux in the
$\upsilon=2\rightarrow0$ band had increased by $\sim2.5$ times during the
following three weeks. 

We note that our model spectrum does not match the observed spectrum at a
number of wavelengths in the SiO band. This is almost certainly due to the
presence of other, as yet unidentified, lines.  Five strong unidentified
emission lines can be seen in the region shortward of the SiO band head; their
observed wavelengths are 3.9907, 3.9926, 4.0003, 4.0026 and 4.0048~$\mu$m. It
is probable that other such lines overlap the SiO band.

\section{Evidence of Infall}

\begin{figure}
\centering
  \includegraphics[angle=-90, width=8cm]{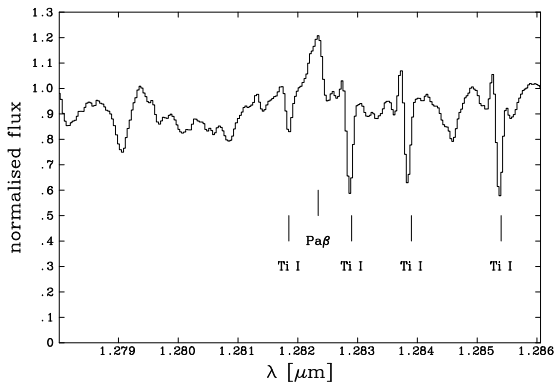}
  \caption{A high resolution spectrum of V838 Mon in the region of Pa$\beta$ on 2003 December 6.}
  \label{highresti}
\end{figure}

Figure~\ref{highresti} shows the high resolution spectrum obtained in the wavelength interval
1.278--1.286~$\mu$m. Clearly visible are a sequence of three spectral lines bearing
well-defined inverse P Cygni profiles with their absorptions centered at 1.2829, 1.2838 and
1.2854~$\mu$m. The only spectral features which can possibly account for these lines are the
permitted transitions of \pion{Ti}{i} at 1.282512, 1.283495 and 1.285054~$\mu$m. These
resolved \pion{Ti}{i} lines are all remarkably narrow, with an average deconvolved FWHM of
$27\pm4$\kms. The cores of the absorptions occur at a heliocentric radial velocity of
$+88\pm7$\kms; the peaks of the emission components are located at $+65\pm6$\kms. All of
these \pion{Ti}{i} lines originate from the b3--z3fo excited states \citep{hoof}. The excitation energies
of the b3 lower levels lie between 1.43 and 1.46~eV above the ground state, and thus could be
collisionally excited in a low temperature gas. There is a further \pion{Ti}{i} line
(1.28115~$\mu$m) present in this spectral region at an observed wavelength of 1.28185~$\mu$m,
but the lower level of this a3D--y3fo transition is further above the ground state (2.16 eV).
This line is in absorption, but may also possess a weak emission component. It is difficult
to identify the continuum level in this spectral region.

\section{Hydrogen emission}

A notable feature of several low resolution $J$ band spectra of V838 Mon that
we have obtained has been the presence of an emission feature near 1.28~$\mu$m
\citep{evans}. This feature is present in the high resolution spectrum shown in
Figure~\ref{highresti}. The most logical identification for it is  Pa$\beta$
(1.282159~$\mu$m). This identification has not been confirmed by detection of
other lines. Pa$\alpha$ (1.875613~$\mu$m) occurs in a region of heavy H$_{2}$O
blanketing in the star and strong attenuation by telluric H$_2$O. Pa$\gamma$
was also not seen in the low resolution data and was not expected given the
weakness of the Pa$\beta$ emission. Br$\alpha$, another expectedly weaker
hydrogen line, is not apparent in the low resolution data presented here in
Figure~\ref{lband}. 

The Pa$\beta$ line peaks at a heliocentric radial velocity of $+56\pm9$\kms.
The red wing of the emission is almost absent. This is a signature of an
accelerating/decelerating flow \citep{bohm}, but the superposition of other
spectral features could also account for the asymmetric profile. Pa$\beta$
absorption, redshifted by $+96\pm6$\kms\, (heliocentric), is a possible
explanation.

\section{Discussion}

We have presented infrared observations of V838 Mon which show SiO overtone emission,
Pa$\beta$ emission and \pion{Ti}{i} lines with inverse P Cygni profiles.

There have not been many reported occurrences of SiO overtone emission in astronomical
objects. Of the few known instances, the phenomenon has been observed in SN 1987A
\citep{meikle}, the Mira variable, $o$ Ceti \citep{yamamura}, the RV Tauri star, R Sct
\citep{matsuura}, and is suspected in the latest M giant stars.

The observed SiO bandheads and band structure of the M giant stars SW Vir and RX Boo have
been shown to be too weak as compared with the predictions by model atmospheres with
appropriate effective temperatures \citep{tsuji94}. Similar results were noted by
\citet{rinsland} and \cite{tsuji}, who suggest that SiO emission occurs in the outer
atmospheres of the coolest O-rich giants, and fills in the strong photospheric SiO absorption
lines.

The presence of inverse P Cygni profiles in the post-eruption spectra of V838 Mon is clear
evidence that some of the expelled material is now falling toward the star. The presence of
SiO and Pa$\beta$ emission  can be understood if this infall is being compressed and
shock-heated by material close to the star, where there are the  high densities necessary to
collisionally excite the SiO (see $\S$3). This is similar to the phenomenon seen in pulsating
variables where rising and falling layers lead to repetitive shock and line emission very
close to the photosphere \citep{matsuura,ferlet,gillet}. 

The SiO emission is redshifted by the same velocity as the \pion{Ti}{i}
absorption and, if present, the Pa$\beta$ absorption. We therefore conclude
that these atomic features must (like the SiO) arise very close to the star.
The  velocity shifts displayed by the \pion{Ti}{i} and Pa$\beta$ emissions are
similar. The simplest explanation is that \pion{Ti}{i} and Pa$\beta$ trace the
same material, with a spherically symmetric distribution, but the difference in
excitation suggests that Pa$\beta$ arises in a deeper layer. Our present
observations do not allow us to determine whether additional velocity
components are present in the SiO emission.

Knowledge of the stellar velocity is required in order to convert our measured
velocities to the rest frame of the star. This parameter is uncertain due to
the presence of spectral lines with P Cygni profiles in the outburst spectra,
and the non-detection of molecular rotational emission \citep{rushtonb}.
Nonetheless \citet{kipper} suggested that the systemic velocity could be
$+59\pm6$\kms\, on the basis of two such velocity components in the P Cygni
emissions in the outburst spectra. {\em If} this is the stellar velocity then
the infall velocity is $\sim15$\kms, but again we emphasize the uncertainty in
the systemic velocity. The many different outflow velocities and components
displayed by the spectral features in the outburst spectra demonstrated the
compexities of the gas motions in V838 Mon at that time.

Titanium is the only atomic species with transitions involving low-lying
states in the high resolution spectral range presented here.
Follow-up observations are necessary to search for infall signatures in other
regions of the spectrum, and to monitor the effect of this infall on the future
behaviour of V838 Mon.

\section*{Acknowledgements} We thank the referee, Geoff Clayton, for his
helpful comments. MTR is supported by a Particle Physics and
Astronomy Research Council (PPARC) studentship. TRG is supported by the Gemini
Observatory, which is operated by the Association of Universities for Research
in Astronomy, Inc., on behalf of the international Gemini partnership of
Argentina, Australia, Brazil, Canada, Chile, the United Kingdom, and the United
States of America.

\bsp

\label{lastpage}

\end{document}